# A method for partitioning the information contained in a protein sequence between its structure and function

Andrea Possenti[1,2], Michele Vendruscolo[2], Carlo Camilloni[3*] and Guido Tiana[1*]

[1]Center for Complexity and Biosystems and Department of Physics, Università degli Studi di Milano and INFN, via Celoria 16, 20133 Milano, Italy

[2]Department of Chemistry, University of Cambridge, Lensfield Road, Cambridge, CB2 1EW, United Kingdom

[3]Dipartimento di Bioscienze, Università degli Studi di Milano, via Celoria 26, 20133 Milano, Italy

**Corresponding authors**:

Guido Tiana, Center for Complexity and Biosystems and Department of Physics, Università degli Studi di Milano and INFN, via Celoria 16, 20133 Milano, Italy, email guido.tiana@unimi.it

Carlo Camilloni, [3]Dipartimento di Bioscienze, Università degli Studi di Milano, via Celoria 26, 20133 Milano, Italy, email: carlo.camilloni@unimi.it

**Abstract**

Proteins employ the information stored in the genetic code and translated into their sequences to carry out well-defined functions in the cellular environment. The possibility to encode for such functions is controlled by the balance between the amount of information supplied by the sequence and that left after that the protein has folded into its structure. We developed a computational algorithm to evaluate the amount of information necessary to specify the protein structure, keeping into account the thermodynamic properties of protein folding. We thus show that the information remaining in the protein sequence after encoding for its structure (the 'information gap') is very close to what needed to encode for its function and interactions. Then, by predicting the information gap directly from the protein sequence, we show that it may be possible to use these insights from information theory to discriminate between ordered and disordered proteins, to identify unknown functions, and to optimize designed proteins sequences.

**Introduction**

Proteins have evolved to perform efficiently their biological functions within a complex environment where they coexist with myriad other molecules[1]. As a consequence, their amino acid sequences include information not only about their structures and functions but also about their interactions with a variety of cellular components, including in particular those required for their homeostasis.

The interpretation of proteins as part of the flow of information from the genetic code to cellular metabolism dates back at least to Francis Crick's central dogma of molecular biology[2]. Anfinsen's experiments[3] clarified one crucial step of this flow - that of going from the protein sequence to its native structure. Since then, information theory has been widely used to investigate protein folding[4-9]. The general strategy is to evaluate the number of bits necessary to describe the native state of a protein and compare it with that encoded by its sequence.

In parallel, in the last several decades, many other aspects of the physics of protein folding have been understood. In particular, it became apparent that proteins are 'frustrated', as many potentially conflicting requirements, including their structures, functions, and the avoidance of misfolding and aggregation, enter in the stabilization of their ground states[10-14]. As a consequence, most proteins are stabilized cooperatively[10,15,16], with a burden that is typically distributed among a minority of residues[17]. In addition, their folding follows a nucleation mechanism driven by a small subset of residues[18-21], in which the correct placing of these few is sufficient to find rapidly the native state.

The physical properties of proteins determine to a large extent the amount of information that should be stored in their sequences. For example, the polypeptide nature of these molecules allows one to describe the position of each triplet of backbone atoms essentially with two dihedrals instead that with nine Cartesian coordinates, thus making use of less sequence-encoded information. This is the core of the approach followed in ref. 5, where for a simple lattice model of proteins, the amount of information needed to specify the native structure was found to be 0.5 bit per residue.

To elucidate quantitatively how proteins can be endowed with these properties, we calculated how the information content in their sequences is partitioned between their structural and functional properties and is thus optimized to confer them the ability to be functional in complex environments.

The goal of the present work is to investigate the flow of information from sequence to structure, and eventually to function of single-domain proteins. For a given protein, we calculated the amount of information $K_{seq}$ [22] provided by its sequence and we compared it with the amount of information $K_{struct}$ needed to define its structure, calculated from the minimum number of physical contacts that define univocally the native structure. With this procedure we built a isomorphism from sequence to structure space, and thus made the associated entropies comparable[23]. Following this strategy, we expressed Anfinsen's hypothesis [3], that the native structure of a protein is completely encoded in its sequence corresponds, as the simple rule: $K_{seq} > K_{struct}$.

We then investigated how proteins exploit the information gap (defined as $\Delta K = K_{seq} - K_{struct}$) between sequence and structure. This gap should be important as proteins, after folding into their native states, should perform their functions and do so in a crowded and regulated environment. Some residues on the surface of the protein should then be allocated as binding sites for other molecules; such residues are very specific, as suggested by the fact that they are highly conserved in orthologous proteins[24], and thus they are expected to require a large amount of information to be carefully specified. Moreover, proteins should fit the cellular homeostasis, and consequently should avoid aggregation and are eventually degraded. These processes require additional information. We define $K_{cell}$ as the total information needed to account for all the requirements associated with the role of a protein in the cell homeostasis.

Our main result is that $K_{seq} \approx K_{struct} + K_{cell}$ (i.e. $\Delta K = K_{cell}$) for all the proteins that we studied, namely that the effect of evolution and obedience to physical laws[25] seems to make the information content of protein sequences perfectly optimized to carry out their biological functions. Furthermore, we showed that it is possible to calculate the information gap of a protein given only its sequence. This calculation allowed us to distinguish ordered and disordered proteins and could in principle be a measure to identify unknown functions in proteins and a further tool for sequence optimization in protein design.

**Methods**

Protein sequences $\{\sigma_i\}$ are stored in the genes of the organism and expressed through their transcription and translation; the amount of information necessary to specify a sequence is given by the self-information $-\log_2 f(\{\sigma_i\})$, where $f(\{\sigma_i\})$ is the *a priori* probability [26] of the sequence $\{\sigma_i\}$, that is its measure in the space of the viable sequences (viable from the point of view of gene encoding, independently on their folding and function). The coding theorem [27] states that for long sequences the self-information approximates the Kolmogorov complexity, that is the minimum number of bits necessary to encode for the sequence.

To quantify the *a priori* probability, one can assume that amino acids are uncorrelated, and thus $f(\{\sigma_i\}) = \prod_i f(\sigma_i)$, where f(σ) is the cellular frequency of the residue of type σ. This assumption follows from the fact that the *a priori* probability only accounts for how sequences are generated from the genome through transcription and translation (as from an Universal Turing Machine [26]), independently of how this information will be used. Indeed, protein sequences can display correlations [28], but these arise *a posteriori* because only some of all the viable sequences are selected by evolution to have specific features (such as to fold to a unique structure), and do not contribute to the *a priori* probability, which only reflects genetic encoding, cellular transcription and translation.

The amount of information associated with the sequence of each protein can thus be written as

$$K_{seq} = -\sum_{i=1}^{N} \log_2 f(\sigma_i). \tag{1}$$

This is analogous to the quantification of the amount of information carried by sequences of letters that form words in natural languages[29], with the important difference that words display correlations *a priori,* for example due to the need of alternating vowels and consonants.

The amount of information needed to encode for the structure of a protein is defined from the minimum amount of (discrete) constraints that confine a protein in its native state, keeping its RMSD below a given threshold. It consists of two contributions, and is defined as

$$K_{str} = K_{ter} + K_{loc}. \qquad (2)$$

$K_{ter}$ is the entropy associated to the tertiary structure of the protein and is given by the sum of a statistical and a multiplicity term as

$$\begin{aligned} K_{ter} &= K_{stat} + K_{mul} \\ &= -\sum_{(\sigma,\tau)=1}^{C} \log_2 p(\sigma,\tau) + \sum_{(\sigma,\tau)=1}^{C} \log_2(n_\sigma n_\tau) \\ &= -\sum_{(\sigma,\tau)=1}^{C} \log_2 \frac{p(\sigma,\tau)}{n_\sigma n_\tau}, \end{aligned} \qquad (3)$$

where C is the minimum number of constraints to be set on the protein three-dimensional arrangement in order to keep it folded, $p(\sigma,\tau)$ is the probability to have a constraint between two residues of type $\sigma$ and $\tau$ within the minimum set of constraints as obtained from contact statistics on the PDB[30] and $n_\sigma$ is the number of residues of type $\sigma$ (the product $n_\sigma n_\tau$ is proportional to the information cost to be paid to select the correct constraint among all possible combinations).

The calculation of the minimal set C of contacts that defines the native state is the most computationally intensive step of the procedure and is obtained with an iterative Monte Carlo algorithm (see Supplementary Information). Several minimization trajectories were computed using the algorithm (see Fig. S1 as an example for the villin headpiece). Although the value C obtained in the minimization is always the same, the individual contacts can be slightly different (typically in hairpins contact *i-j* can be substituted by *i-(j±1)*, see Table S1). Since $K_{ter}$ depends also on the kind of residues in the minimal set (cf. Eq. (3)), this gives some variability in the estimate of $K_{ter}$. For example, in the case of villin we obtained $\overline{K_{ter}} = 111.34 \pm 5.6\ bits$ (see caption of Table S1), corresponding to a relative error of ≈5%.

The second contribution to the structural algorithmic entropy, the one associated with the secondary structure $K_{loc}$ is defined as

$$K_{loc} = -\sum_{i=1}^{C_{loc}} \log_2 p_{loc}^{(i)}, \tag{4}$$

where $C_{loc}$ is the total number of local constraints imposed to the protein (see text and Fig. S2 in the Supplementary Material), and $p_{loc}^{(i)}$ is the probability of each residue belonging to the sequence to populate a given secondary structure and is computed using PSIPRED software [31].

In general, the algorithmic entropy associated with two different spaces cannot be compared with each other. However, in the present case the fact that the map between the sequence and the structure goes through restrains that can be interpreted as physical interactions builds an isomorphism from sequence to structure space, allowing the sequence entropy to be compared with the structural one[23].

Cells use the amount of information contained in the sequence to produce proteins with the needed features in terms not only of equilibrium structure, but also of presence of interaction with the cellular environment. In this respect, evolution selects and encodes in the genome only the subset of those sequences that display useful features for its functioning (using again the analogy of natural languages, only "meaningful" words).

The amount of information contained in the sequence is then used to make the protein functional in the cellular environment. We quantified the amount of information needed for this purpose as $K_{cell}$, given by the sum of three contributions

$$K_{cell} = K_{cleav} + K_{solub} + K_{fun}. \tag{5}$$

Here, $K_{cleav}$ contains the information related to the probability of each residue to be the site of the proteolytic cleavage and is defined as

$$K_{cleav} = -\sum_{i=1}^{N} \log_2 p_{cleav}^{(i)}, \tag{6}$$

where $p_{cleav}^{(i)}$ is the cleavage probability for each residue, and it is computed using MAPPP software [32] (for further detail see text in the Supplementary Material).

The second term in eq. (5) $K_{solub}$ contains the information associated to the solubility and aggregation propensity of the protein and is given by

$$K_{solub} = - \sum_{i=1}^{N} \log_2 p_{solub}^{(i)}, \tag{7}$$

where $p_{solub}^{(i)}$ is the probability associated to each residue along the chain to be considered as soluble. The solubility profile is initially computed using CamSol software [33], and the probability is obtained as the score of each soluble residue normalised over the maximum score of the profile.

The last term $K_{fun}$ in eq. (5) accounts for the biological function of each protein studied. It's the information content related to the protein binding sites, obtained through a literature search, and is defined as

$$K_{fun} = - \sum_{i=1}^{N_{bind}} \log_2 p_{bind}(\sigma_i), \tag{8}$$

where $N_{bind}$ is the total number of binding sites and $p_{bind}(\sigma_i)$ is the probability of the binding site to be the residue of type $\sigma$.

An important approximation associated with Eq. (5) is that that proteolytic cleavage, solubility and binding site are assumed as independent on each other. As discussed in the Results Section below, the overlap between sites involved in the different terms in which we partitioned entropy are, in general, not statistically significant.

**Results**

We calculated the amount of information needed for a protein sequence to encode most, if not all, the relevant information needed to be functional in an environment that, although described in a simplified way, retains the main cellular features.

*Sequence entropy.* The amount of information required to specify the sequence of a given protein of length $N$ out of the $20^N$ possible amino acid sequences is quantified by the algorithmic entropy $K_{seq}$ (**Fig. 1**). $K_{seq}$ can be estimated from the frequencies of the 20 types of amino acids in proteins (see Materials and Methods and **Table S2**), exploiting the observation that they are essentially uncorrelated in protein sequences[34]. $K_{seq}$ is thus the information available in a protein sequence to encode all its properties.

*Structure entropy.* The amount of information needed to encode the structural features cannot be determined neglecting the cooperative nature of protein folding. If the structural information were calculated as the amount needed to constrain all the dihedrals of a protein, independently on each other, to define the native state within a given resolution (e.g. 4 Å), similarly to what was done in ref. [5], would require an amount of information larger than that contained in the sequence. For example, this calculation would give for the native structure of B1 immunoglobulin-binding domain (pdb code 1PGB) an amount of information $K_{struct}$ to specify the structure of 363 bit (see caption of **Fig. S2**), which is more than 50% larger than the information available from its sequence ($K_{seq}$ of 234 bit). This discrepancy would indicate that the sequence alone cannot determine the structure, and thus would not agree with Anfinsen's thermodynamic hypothesis.

Keeping into account the nucleated mechanism of protein folding, the value of $K_{struct}$ is much reduced. We calculated $K_{struct}$ as the sum of a term $K_{loc}$, accounting for the algorithmic entropy needed to constrain the chain to assume native secondary structures (α-helices and β-strands, **Table S3**), and a term $K_{ter}$ associated with the formation of the tertiary structure. The latter is calculated from the minimum number of contact constrains (**Figs. 1 and 2**) which assures that the protein is in the native state within a given resolution. The algorithm to calculate $K_{ter}$ requires the knowledge of the native conformation of the protein and is computationally quite demanding, scaling as $N^3 log\ N$ with the length $N$ of the protein (see Materials and Methods).

Both the local and the tertiary constraints are implemented in a way to mimic the physical interactions between residues, to reflect the very physical mechanism employed by the sequence to encode the structure. In this way, the isomorphism between sequence and conformational space needed to make the associated entropies comparable [23] is realized by the physical mechanism of molecular interaction.

The calculated values of $K_{struct}$, $K_{loc}$ and $K_{ter}$ for five structured proteins of different lengths are reported in **Fig. 1** (see also **Table S4**). In all cases we found that the amount of information $K_{struct}$ needed to specify the structure of a protein is less than that, $K_{seq}$, provided by its sequence, in agreement with Anfinsen's thermodynamic hypothesis. From **Figs. 1** and **2** it is apparent that the information needed to specify the local structure of the chain, contributing to the definition of its secondary structure is a small fraction of the total structural information, being of the order of 5-10% and reaching 25% only for lysozyme (pdb 2LYZ).

Furthermore, a small number of tertiary contacts (**Table S5**) is enough to determine, together with the former local constraints, the global structure of the 5 proteins. Interestingly, these pairs of tertiary contacts form a network of key interactions (**Figs. 2** and **S3**) displaying a large diameter, comparable with that of the whole network of native contacts, and small transitivity (**Fig. S4** and **Table S6**). Thus, the network of key interactions forms a sort of irregular lattice [12], which is particularly economic to define the native structure.

The sites involved in the key interactions are not particularly conserved among sequences displaying the same native structure, compatibly with the idea that evolution could find different solutions to stabilize a protein fold [35]. However, the corresponding pairs display a significantly large amount of direct information (**Table S8** and **Fig. S5**), meaning that the mutations in these sites tend to coevolve to maintain the stability of the native structure [28]. Particularly interesting from this perspective is 2LYZ, for which the algorithm found two key contacts corresponding to the disulfide bridges which lock the native conformation (**Fig. 2**).

*Cellular entropy.* In accordance with the sequence – structure – function paradigm, the observed excess of information in protein sequence with respect to protein structure, $K_{seq} > K_{struct}$, could then be used to encode for other features. In the global context of cell homeostasis, proteins are evolved to be functional that also means soluble and disposable [36]. We thus quantified the amount of information needed to specify the solubility of the proteins, the presence of proteolytic sites to control their metabolism in the cell, and the presence of active sites involved in catalytic activity or binding (see Materials and Methods). The overall functional information is then here termed $K_{cell}$. All these quantities are reported in **Fig. 3** and compared with the size $K_{seq} - K_{struct}$ of the information gap between sequence and structure. Interestingly, in all cases they fill almost exactly the information gap, that is $K_{cell} \approx K_{seq} - K_{struct}$ (see also **Table S9**). In **Fig. 2** is displayed how the information provided by the sequence is shared between the residues of the proteins to carry out the different tasks (functional sites, cleavage sites, solubility sites, sites responsible for local or tertiary structure). The overlap between the different roles is usually low and is statistically highly significant between cleavage and solubility sites, between functional and solubility sites and between cleavage sites and those responsible for tertiary structure (see horizontal bars in **Fig. 3** and **Table S10**). The histograms displayed in **Fig. 2** indicate the amount of information necessary to specify each role. Not unexpectedly, the information needed to specify tertiary structure, cleavage and, to some extent, solubility, is spread homogeneously over the sites of all studied proteins. That associated with function is not and, remarkably, also that associated with secondary structures is not, although secondary structures are distributed rather evenly in these proteins.

A striking feature that we observed (**Fig. 3**) is that both $K_{seq}$ and $K_{struct}$, and consequently the information gap, increase linearly with the size of the protein up to ~150 residues, which is the approximately the average size of single-domain proteins in eukaryotes [37]. The amount of information required for solubility and cleavage is also approximately linear with protein size (see also **Table S9**), and consequently that left for activity also increases linearly with size. The linear behavior observed in **Fig. 3** for $K_{struct}$ can be then in principle used to predict the information gap of proteins from their sequence, without any complex structural calculation, assuming the linear fit

$$K_{struct} = -6.32 + 3.22 \text{ N}. \tag{9}$$

To test this hypothesis, we calculated $K_{seq} - K_{sol} - K_{cleav}$ for 24 proteins of various length (see also **Table S11**), chosen with the only requirement of not being listed in the DisProt database [38], and thus be, presumably, structured. The yellow and the purple dots in **Fig. 4a** indicate the value of $K_{seq}$ and $K_{seq} - K_{sol} - K_{cleav}$, that is the amount of information which is available for structure and for function. In most cases, the purple dots lie well above the straight line which marks the fit of $K_{struct}$ (that is Eq. (9)), indicating that these proteins have information available for displaying specific functions. Exception is made for two proteins (Q9C010 and P56211, see **Table S11**), which apparently do not display any usable gap. However, the analysis of these two proteins with MobiDB [39] and s2D [40], show that they are actually disordered proteins. Consequently, the corresponding points can lie below $K_{struct}$ simply because they do not need that amount of information to become structured.

*Disordered proteins.* This finding suggests another possible use of Eq. (9), which is to look at the behavior of the information content in the case of intrinsically disordered proteins (IDPs). In **Fig. 4b** we analyzed the degree of information of 9 IDPs obtained from DisProt and confirmed by s2D; the associated value of $K_{seq} - K_{sol} - K_{cleav}$ are there with purple triangles. Strikingly all of them are below, or anyway within the error bars (gray line) of the linear fit of $K_{struct}$. This result suggests the fascinating hypothesis that IDPs have much greater freedom than ordered proteins to modulate their information content toward functions in place of structure. Such greater freedom is consistent with the observation that IDPs tend to have multiple partners, as required by their key roles in signaling and regulation.

*Designed proteins.* Finally, one can wonder if designed proteins shown a different signature with respect to naturally evolved ones. We thus studied the information content of rationally designed proteins from two recent studies [41,42]. Interestingly, as shown in **Fig. 5**, all of them, with the exception of 1 protein from the older set (see **Table S12**), are within the error bars or well above the linear fit of the structural information. The presence of one designed protein in the "IDPs region" could be explained by analyzing its solubility profile (cf. Fig. S9). This protein looks highly soluble across the whole sequence, which in our context means that a great part of the information is assigned to the solubility, and that, in a broader biological perspective, this could affect its structural stability. This is also interesting with respect to the observation in [41,42] that improve in hydrophobic interactions was key to improve the design. Indeed, upon comparison between the two different designs, it can be noticed how the most recent iterative design approach (red diamonds in Fig. 5) appears to give better results in terms of availability of structural information than the previous one (green diamonds).

**Discussion**

The genetic code stored in the DNA contains most of the information needed by proteins to generate and maintain cellular life. Part of this information is translated into protein sequences and can be readily quantified using information theory [4-9]. In turn, the information contained in protein sequences can be exploited to carry out function, through specific binding or catalytic sites, and regulation, hosting sites which permit proteolysis and avoid aggregation [43]. The information contents needed for these functions can be quantified as well. Protein function is usually mediated by structural features, and we could quantify the amount of information necessary to generate local secondary structure and, for structured proteins, tertiary structure. The latter is encoded in the sequence in a complex way. In particular, the necessity that the protein obeys the laws of polymer physics makes the evaluation of $K_{struct}$ computationally uneasy, but reduces largely the amount of information needed for its determination. In fact, we have shown that a way to verify quantitatively Anfinsen's thermodynamic hypothesis is to keep into account the cooperativity of protein folding associated with its nucleation character (i.e. the need of few key interactions).

The estimates we presented here for some structured proteins indicate that the amount of information stored in the sequence is approximately equal to that needed by structure and cellular functions. Although in the present treatment we have neglected the additional information provided through other cellular processes, including post-translational modifications and sub-cellular localization, this result suggests that structured proteins are highly optimized, exploiting essentially all the information they have from the genetic codes, and that there is little room for the addition of new functions without major changes in the protein [44]. The width of the information gap available for cellular functions increases linearly with protein size. Thus, in principle, proteins experience evolutionary pressure towards larger sizes to increase their functionality. The evidence that single-domain structured proteins reaches up to lengths of ~150 residues [37] and larger proteins are multi-domain suggests a physical barrier against this tendency, most likely associated with the impossibility of maintain solubility [45]. Within this framework, we observed that the functional contributions systematically fill the information gap between sequence and structure regardless of the protein length (see **Fig. 3)**. This is consistent with the idea that proteins' intrinsic properties have been finely tuned by evolution to optimize protein functionality, solubility and turnover within the cell, in agreement with the hypothesis that proteins are expressed in the cellular environment at critical levels which maintain them functional, yet still soluble [46].

From our estimates, it is clear that most of the available information is used to build the tertiary structure, with only a small fraction devoted to stabilize secondary structure. Here IDPs can take advantage of the saved information, i.e. the lack of stable tertiary structure, to expand their functionality, without the need to introduce any special framework [47]. This result is in agreement with the accepted idea that IDPs tend to be more promiscuous than structural protein in protein-protein interactions [47,48].

From the evolutionary point of view, the existence of a very small gap in structured proteins between the amount of information provided by the sequence and that used by the protein is compatible with some key observations. For example, the coevolution of residues that stabilize tertiary structures [28] is what one expects if along evolution the protein is constrained in changing the stabilizing pairs, not to increase $K_{ter}$. Similarly, the tiny gap is compatible with the limited evolutionary divergence of proteins that perform the same function, and with the fact that structural divergence usually implies functional divergence but not vice versa, because the amount of bits necessary to specify the former is 1-2 orders of magnitude larger than the latter [49].

The existence of a simple, general expression for $K_{struct}$ as a function of protein length allows one to use the amount of information as a predictive tool. Since $K_{sol}$ and $K_{cleav}$ can be estimated easily, it is rather straightforward to estimate the number of sites which can potentially host binding or catalytic function. Moreover, the difference observed in **Fig. 4** between structured proteins and IDPs can be used as predictor of disorder and/or as a score for unknown functions in the direction of developing new measures for proteins design (see **Fig. 5**).

Overall, we have used information theory to investigate the consequences of the evolutionary balance between mutation and selection in proteins, which generates the information stored in their amino acid sequences. We found that this balance results in an almost perfect partitioning between structure and function of the information stored in the sequences themselves, which can be exploited to obtain relevant insights into the behavior of these molecules.

FIGURES

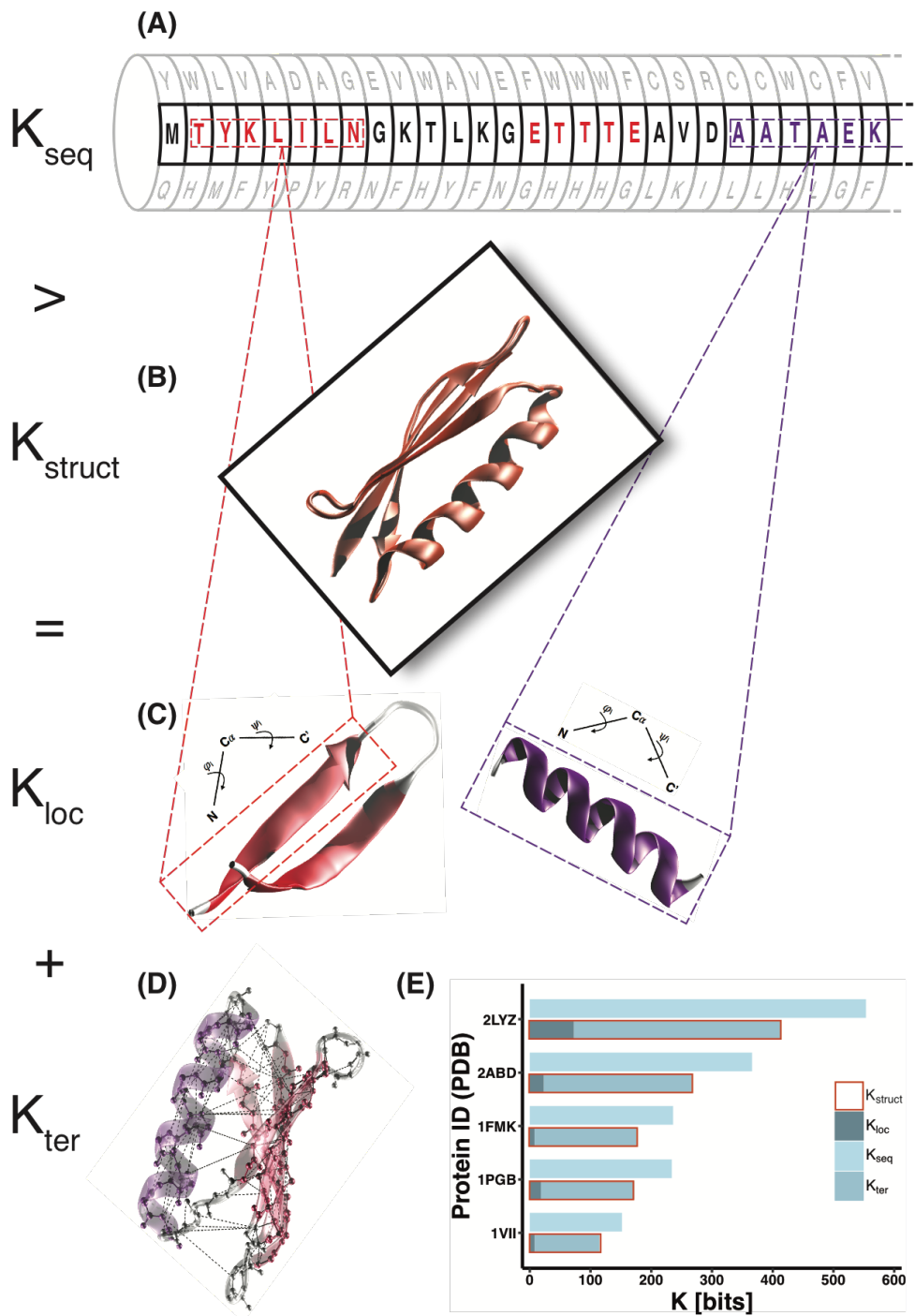

**Figure 1**. **The information content of proteins sequences is larger than that needed for proteins structures.** **(A)** The primary structure of a protein of N amino acids is one out of $20^N$ possible realizations, where 20 is the number of amino acids. The quantity $K_{seq}$ is the amount of information needed to specify it. **(B)** The overall information $K_{struct}$ needed to specify the structure is the sum of the two contributions, $K_{loc}$ and $K_{ter}$: **(C)** $K_{loc}$ that is the amount of information needed to specify the local arrangement of the secondary structures, restraining its Ramachandran dihedrals, **(D)** $K_{ter}$ that is the information associated with the tertiary structure and depends on the minimum number of interactions needed to constrain the protein in its native conformation (dashed lines). **(E)** The comparison between $K_{seq}$ and $K_{struct}$ for five widely-studied proteins, the villin headpiece (1VII), B1 immunoglobulin-binding domain (1PGB), src SH3 (1FMK), ACBP (2ABD) and hen-egg lysozyme (2LYZ).

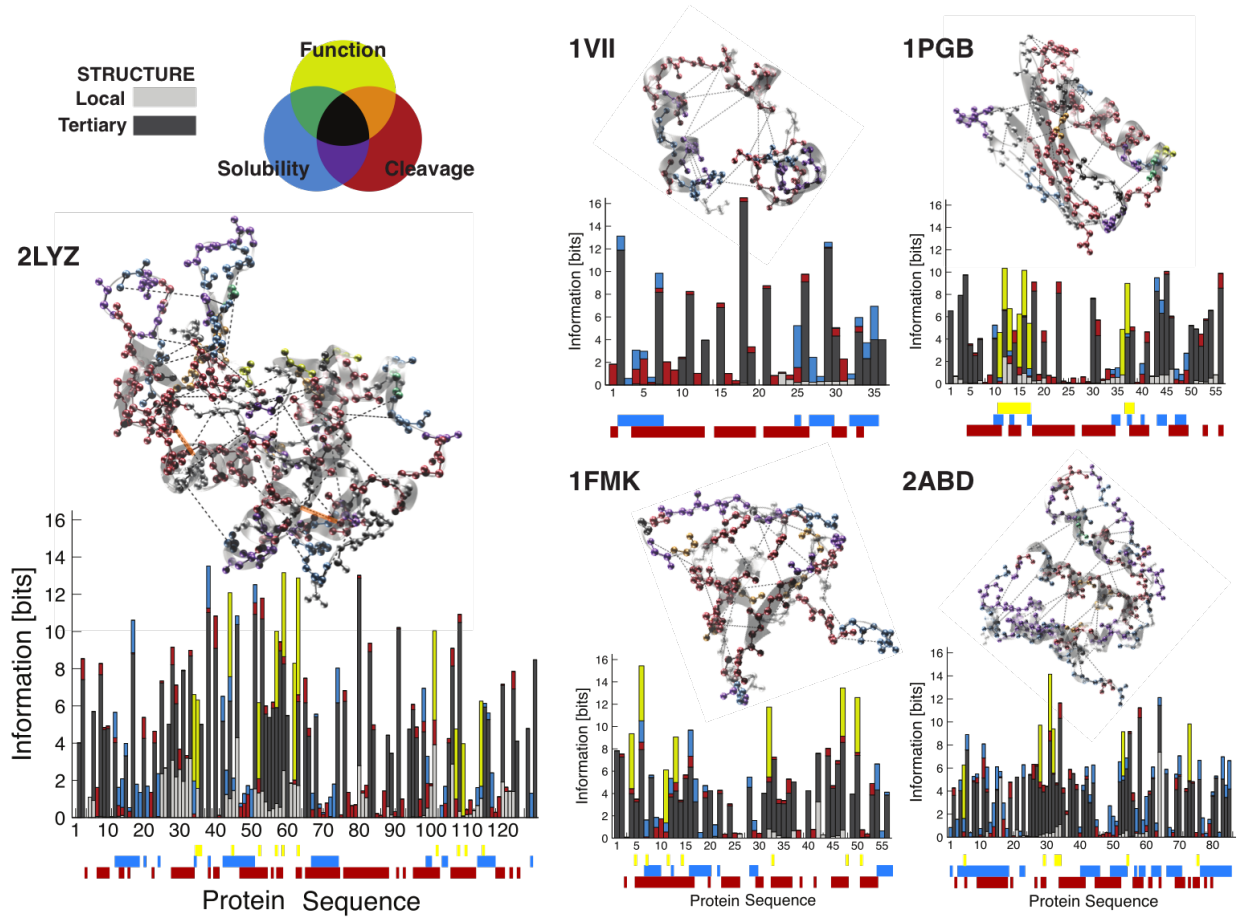

**Figure 2**. **The structural and functional information is unevenly distributed along the sequence.** For the native structure of the five proteins under study the local structures, which define $K_{loc}$, are highlighted with a cartoon while the constraints on the tertiary structure, which define $K_{ter}$, are indicated with red dashed lines. Atoms of each amino acid are colored according to their role (solubility, cleavage and function). The disulfide bonds of lysozyme are marked with orange lines. The same information is reported with colored stripes below each histogram. The histograms report the amount of information one needs to pay to assign to each amino acid its role.

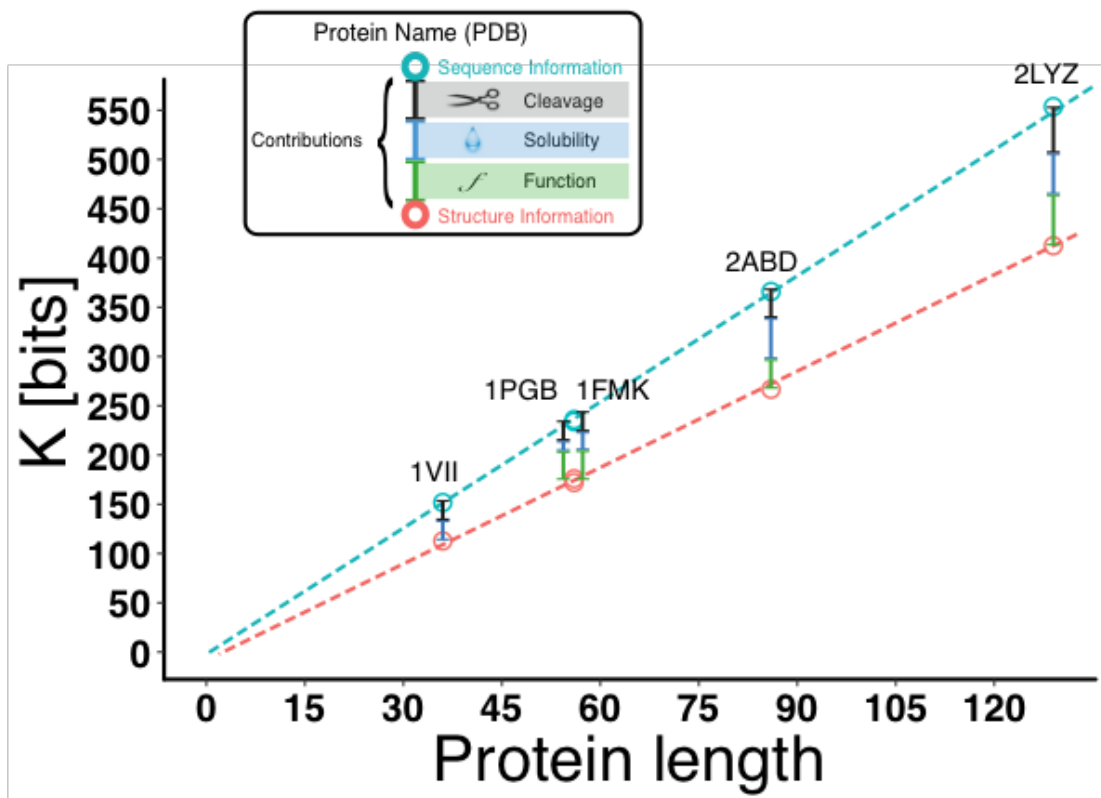

**Figure 3**. **The gap between the information content of proteins sequences and structures can encode for their functions.** The amount of information $K$ provided by the sequence, $K_{seq}$ (red circles) and that used to define the native state, $K_{struct}$ (cyan circles) as a function of its length for five protein domains of increasing length, labeled with their PDB code. The dashed lines indicate the linear fit. The bars indicate the amount of information employed for defining cleavage sites, $K_{cleav}$, using MAPPP (47) (in black), solubility sites, $K_{sol}$, defined using CamSol (48) (in blue) and protein function, $K_{fun}$, as from Table S9 (in green). While the $K_{seq}$ is linear by definition, $K_{struct}$ seems to increase linearly with the length of the sequence with a slope smaller than that of $K_{seq}$. The gap between the two thus increases linearly as well with protein length and is remarkably comparable to the sum of the contribution from cleavage, solubility and function.

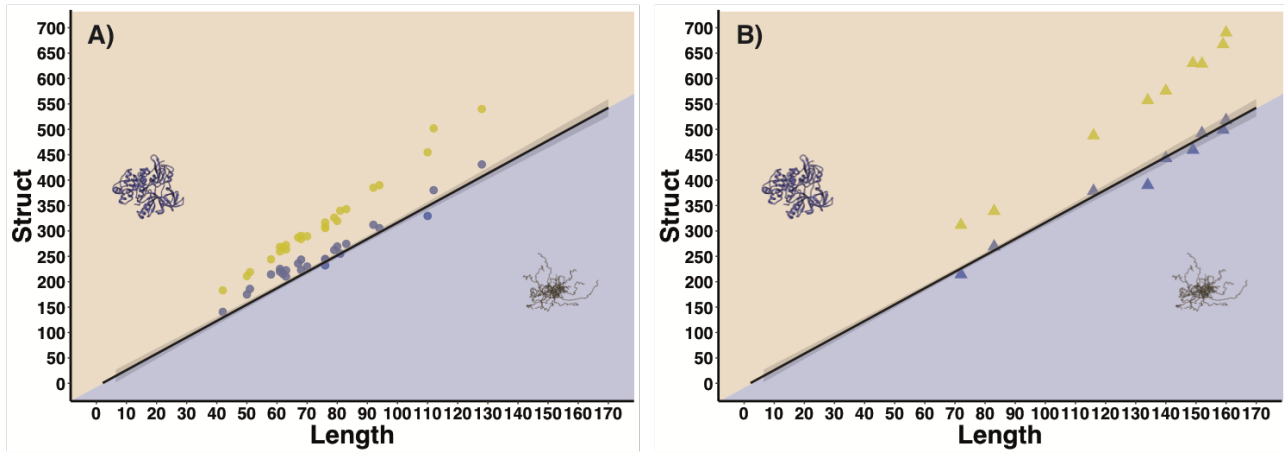

**Figure 4**. **The linear relationship between structural information and protein length can be used to study protein function and structural disorder.** A) Test set: Yellow dots indicate the value of $K_{seq}$ for 24 proteins of variable length. The purple dots indicate the amount of information $K_{seq} - K_{sol} - K_{cleav}$ that can be used for structure and for function. The straight line indicates the values of $K_{struct}$ extrapolated from Fig. 2 and the gray shadow the associated standard error. For 22 out of 24 proteins the difference is above the linear fit suggesting that there is still information available to encode for the function. Two sequences fall below the fit and are indeed found to corresponds to intrinsically disordered proteins. B) Intrinsically disordered proteins set: Yellow triangles indicate the value of $K_{seq}$ for 9 disordered proteins of variable length. Purple triangles indicate $K_{seq} - K_{sol} - K_{cleav}$. The straight line indicates the values of $K_{struct}$ extrapolated from Fig. 2 and the gray shadow the associated standard error. In all cases the IDPs fall below or within the error of the fit suggesting that by subtracting the information needed for solubility and proteolysis there is not enough information left to encode a three-dimensional structure; indeed, all the information left is actually available for local structure and function.

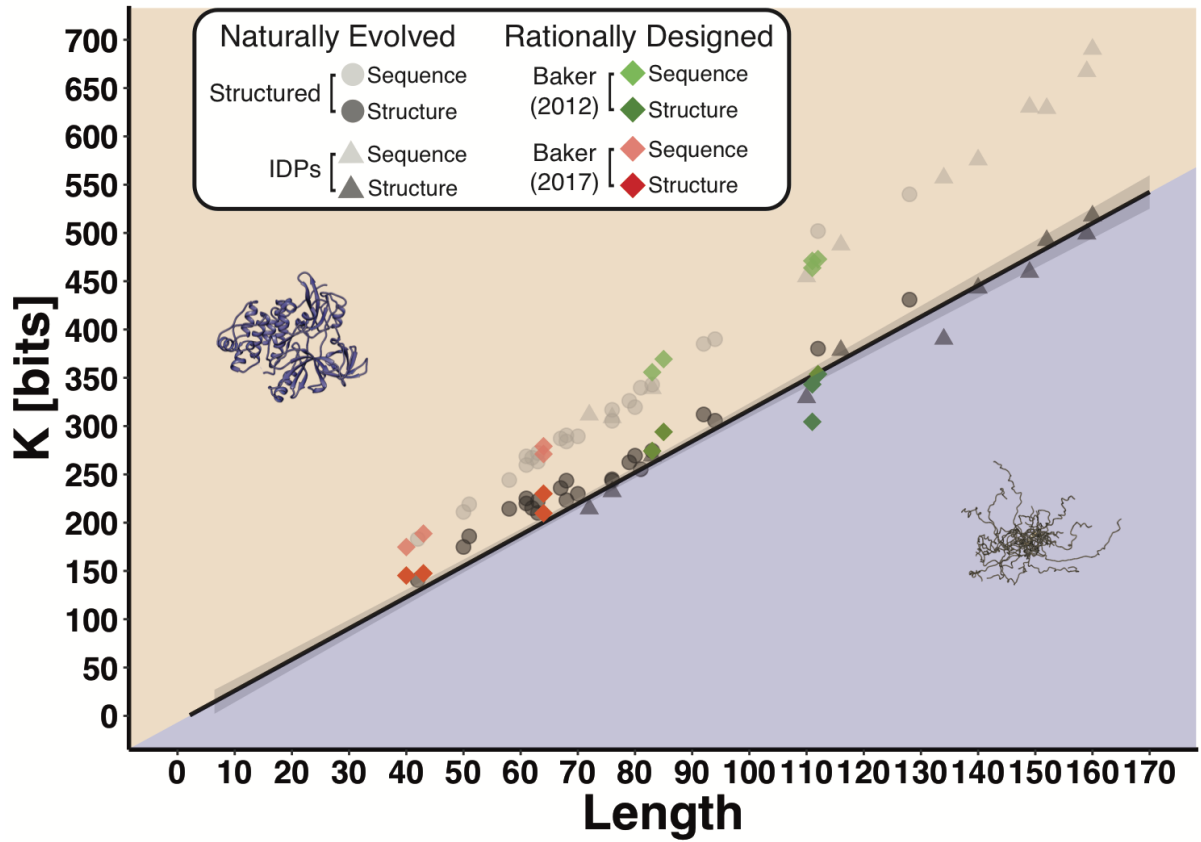

**Figure 5. Comparison between the information content of naturally evolved and rationally designed proteins.** All the designed proteins (solid colors), with one exception (pdb entry 2LTA) are well above - or within the errors - the linear fit. For comparisons are also reported the data from Fig. 4 (transparent colors), where the two disordered proteins from the test set of Fig. 4A are now reported as IDPs (triangles).